\documentclass [12pt,a4paper]{article}
\usepackage{amssymb, theorem}
\newtheorem{thm}{Theorem}
\newtheorem{prop}{Proposition}
\newtheorem{cor}{Corollary}
\newtheorem{lemma}{Lemma}
\theorembodyfont{\rm}
\newtheorem{pl}{Example}

\def\proof{{\it Proof: }}

\def\qed{\nobreak\hfill $\square$}
\def\<{\langle}
\def\>{\rangle}

\def\iA{{\cal A}}

\def\iE{{\cal E}}

\def\Mn{M_n(\bbbc)}

\def\dim{{\rm dim}}
\def\im{{\rm i}}
\def\ot{\otimes}

\def\bbbr{{\mathbb R}}
\def\bbbc{{\mathbb C}}

\def\Tr{\mathrm{Tr}\,}
\def\pont{{\, \cdot \,}}

\def\Mn{M_n}
\begin{document}
\ \vskip 1cm 
\centerline{\LARGE {\bf Generalizations of Pauli channels}} 
\bigskip \bigskip
\centerline{\large D\'enes Petz$^1$ and Hiromichi Ohno$^2$}
\bigskip
\begin{center}
{$^1$Alfr\'ed R\'enyi Institute of Mathematics, H-1364 Budapest,
POB 127, Hungary}\\ \medskip
{$^2$Graduate School of Mathematics, Kyushu University, Japan}
\end{center}
\bigskip
\bigskip\bigskip
\begin{abstract}
The Pauli channel acting on $2 \times 2$ matrices is generalized to an 
$n$-level quantum system. When the full matrix algebra $M_n$ is decomposed 
into pairwise complementary subalgebras, then trace-preserving linear 
mappings $M_n \to M_n$ are constructed such that the restriction to 
the  subalgebras are depolarizing channels. The result is the necessary 
and sufficient condition of complete positivity. The main examples appear 
on bipartite systems. 

\medskip\noindent 
2000 {\sl Mathematics Subject Classification.} 
Primary 15A30, 94A40; Secondary 47N50.

\medskip\noindent
{\it Key words and phrases:}
Completely positive mappings, complementary subalgebras, channel, 
conditional expectation.
\end{abstract}

In this paper we consider particular subalgebras of a full matrix 
algebra $M_n = M_n(\bbbc)$. (By a subalgebra we mean *-subalgebra 
with unit.)  An F-subalgebra is a subalgebra isomorphic to a full 
matrix algebra $M_k$.  (``F'' is the abbreviation of ``factor", the 
center of such a subalgebra is minimal, $\bbbc I$.) An M-subalgebra 
is a maximal Abelian subalgebra, equivalently, it is isomorphic to 
$\bbbc^n$. (``M'' is an abbreviation of ``MASA'', the center is 
maximal, it is the whole subalgebra.) If $\iA_1$ and $\iA_2$ are 
subalgebras, then they are called quasi-orthogonal or complementary 
if the subspaces $\iA_1 \ominus \bbbc I$ and  $\iA_2 \ominus \bbbc I$ 
are orthogonal (with respect to the Hilbert-Schmidt inner product 
$\<A,B\>=\Tr A^*B$\,). Concerning  complementary subalgebras we refer 
to \cite{PDcomp}, see also \cite{Ohno, OPSz, m4}.

Complementary M-subalgebras can be given by mutually 
unbiased bases. Assume that $\xi_1,\xi_2,\dots ,\xi_n $ and $\eta_1,\eta_2,
\dots ,\eta_n$ are orthonormal bases such that
$$
|\<\xi_i,\eta_j\>|=\frac{1}{\sqrt{n}} \qquad (1 \le i,j \le n).
$$
If $\iA_1$ is the algebra of all operators with diagonal matrix in
the first basis and $\iA_2$ is defined similarly with respect to the
second basis, then $\iA_1$ and $\iA_2$ are complementary M-subalgebras.

There are examples such that $M_n$ is the linear span of pairwise
complementary subalgebras in the case when $n$ is a power of a prime number. 
If $M_n$ is decomposed into complementary subalgebras, then we construct 
trace-preserving mappings $M_n \to M_n$ which are completely positive 
under some conditions.

\section{Introduction}

If the pairwise complementary subalgebras $\iA_1,\iA_2,\dots, \iA_{r}$ 
of $M_n$ are given and they linearly span the whole algebra $M_n$, then 
any operator is the sum of the components in the subspaces $\iA_i \ominus  
\bbbc I$ ($1 \le i \le r$) and $\bbbc I$:
$$
A= - \frac{(r-1)\Tr A}{n}I +\sum_{i=1}^r E_i(A)\, , 
$$
where $E_i:M_n \to \iA_i$ is the trace-preserving conditional expectation 
(which is nothing else but the orthogonal projection with respect to the 
Hilbert-Schmidt inner product, see \cite{pd2} about details). It is easier 
to formulate things for matrices of trace 0. If $\Tr B=0$, then it has 
orthogonal decomposition
$$
B=\sum_{i=1}^r E_i(B)\,.
$$
As a generalization of the Pauli channel on a qubit, we define a linear 
mapping $\alpha:\Mn \to \Mn$ such that
$$
\alpha (B) =\sum_{i=1}^r 
\lambda_i E_i(B)
$$
or for an arbitrary $A$
\begin{equation}\label{E:al}
\alpha(A)=\Big(1-\sum_{i=1}^r\lambda_i \Big) \frac{\Tr A}{n}I+ 
\sum_{i=1}^r \lambda_i E_i(A),
\end{equation}
where $\lambda_i \in \bbbr$, $1 \le i \le r$. We want to find the 
condition for complete positivity. The motivation is the following 
well-known example in which the complementary subalgebras are generated by the Pauli matrices \cite{pd2}. 

\begin{pl}\label{2pl1.8}
Let $\sigma_0 =I$ and $\sigma_1, \sigma_2, \sigma_3$ be Pauli matrices, i.e.,
\[
\sigma_1 = \left[ \begin{array}{cc} 0&1\\ 
1&0\end{array} \right], \quad
\sigma_2 = \left[ \begin{array}{cc} 0&-\im\\ \im&0
\end{array} \right], \quad
\sigma_3 = \left[ \begin{array}{cc}1&0\\0&-1\end{array} 
\right]
\]
and let $\iE:M_2 \to M_2$ be defined as
\begin{equation}\label{E:3}
\iE  \left(w_0 \sigma_0 + (w_1,w_2,w_3) \cdot \sigma \right) 
= w_0 \sigma_0 +(\lambda_1 w_1,
\lambda_2 w_2, \lambda_3 w_3) \cdot \sigma
\end{equation}
for $\omega_i \in {\mathbb C}$, where $\lambda_i \in \bbbr$ and
$$
(w_1,w_2,w_3)\cdot \sigma=w_1 \sigma_1 + w_2 \sigma_2 + w_3 \sigma_3.
$$
Density matrices are sent to density matrices if 
and only if
$$
-1 \le \lambda_i \le 1.
$$
It is not difficult to compute the representing block matrix
$X:=\sum_{i,j} \iE(E_{ij})\ot E_{ij}$, we have
$$
X= \left[\matrix{ \frac{1+\lambda_3}{2} & 0 & 0 & \frac{\lambda_1
+\lambda_2}{2}\cr 0 & \frac{1-\lambda_3}{2} & \frac{\lambda_1-
\lambda_2}{2} & 0 \cr 0 & \frac{\lambda_1-\lambda_2}{2} & 
\frac{1-\lambda_3}{2} & 0 \cr\frac{\lambda_1+\lambda_2}{2} 
& 0 & 0 & \frac{1+\lambda_3}{2}}\right].
$$
According to Choi's theorem \cite{Choi} the positivity of this matrix 
is equivalent to the complete positivity of $\iE$. $X$ is unitarily 
equivalent to the matrix
$$
\left[\matrix{\frac{1+\lambda_3}{2} & \frac{\lambda_1+\lambda_2}{2} 
& 0 & 0 \cr\frac{\lambda_1+\lambda_2}{2} &  \frac{1+\lambda_3}{2} 
& 0 & 0 \cr 0 & 0 & \frac{1-\lambda_3}{2} & \frac{\lambda_1-\lambda_2}{2} 
\cr 0 & 0 & \frac{\lambda_1-\lambda_2}{2} & \frac{1-\lambda_3}{2}}\right].
$$
This matrix is obviously positive if and only if
\begin{equation} \label{E:4}
1\pm \lambda_3 \ge |\lambda_1\pm\lambda_2|.
\end{equation}
This is necessary and sufficient condition of complete positivity. \qed
\end{pl}

It is not obvious that condition (\ref{E:4}) is symmetric in the three 
variables $\lambda_1, \lambda_2, \lambda_3$. Condition (\ref{E:4}) actually
determines the tetrahedron which is the convex hull of the points
$(1,1,1)$, $(1,-1,-1)$, $(-1,1,-1)$ and $(-1,-1,1)$.

Now we show the idea leading to the generalization. The mapping $\iE$ in  
(\ref{E:3}) has the form
$$
\iE(\pont)=\sum_{i=0}^3 \mu_i \sigma_i (\pont) \sigma_i.
$$
{F}rom the expansion of $\iE(\sigma_j)$ we can get equations and 
the solution is the following:
\begin{eqnarray*}
\mu_0 &=&\frac{1}{4}(1+\lambda_1 +\lambda_2 +\lambda_3 ), 
\qquad\mu_1=\frac{1}{4}(1+\lambda_1 -\lambda_2 -\lambda_3 ),
\cr\mu_2&=&\frac{1}{4}(1-\lambda_1 +\lambda_2 -\lambda_3 ),
\qquad\mu_3 =\frac{1}{4}(1-\lambda_1 -\lambda_2 +\lambda_3 ).
\end{eqnarray*}
If $\mu_i \ge 0$ for every $i$, then $\iE$ is a completely positive mapping. 
Therefore, 
$$
1+\lambda_3 \ge \pm (\lambda_1 +\lambda_2), \qquad 1-\lambda_3 \ge 
\pm(\lambda_1 -\lambda_2 )
$$
or together this is (\ref{E:4}). (Actually, this argument gives that  
(\ref{E:4}) is a sufficient condition for the complete positivity.)

Pauli channels form an important and popular subject in quantum 
information theory \cite{BFS, Fuji, King}. The mappings (\ref{E:al}) 
were studied in the paper \cite{NR} in the case when the subalgebras 
are maximal Abelian and pairwise complementary. Our method is different 
and we allow non-commutative subalgebras as well.

The mapping (\ref{E:al}) restricted to $\iA_i$ has the form
$$
D \mapsto \lambda_i D + (1-\lambda_i ) \frac{I}{n}
$$
on density matrices $D$. If $ 0 \le \lambda_i \le 1$, then we can say 
that $D$ does not change with probability $\lambda_i$ and  with 
probability $1-\lambda_i$ 
it is sent to the tracial state. Such mappings are usually called as 
depolarizing channels \cite{pd2}.
 
A simple example including non-commutative subalgebras is the following.

\begin{pl}\label{Ex:2}
Consider $M_4=M_2\ot M_2$ and the complementary F-subalgebras $\iA_1,\dots, 
\iA_4$ generated by the following triplets of unitaries:
$$
\begin{array}{ccc}\sigma_0\otimes\sigma_1 & \sigma_0\otimes\sigma_2 & 
\sigma_0\otimes\sigma_3, \\\sigma_1\otimes\sigma_0& \sigma_2\otimes
\sigma_1 & \sigma_3\otimes\sigma_1, \\  \sigma_2\otimes\sigma_0 &
\sigma_3\otimes\sigma_2 & \sigma_1\otimes\sigma_2, \\ \sigma_3
\otimes\sigma_0 & \sigma_1\otimes\sigma_3 & \sigma_2\otimes\sigma_3.
\end{array}
$$
We take also the M-subalgebra $\iA_5$ generated by $\sigma_1\otimes\sigma_1, 
\sigma_2\otimes\sigma_2, \sigma_3\otimes\sigma_3$. The conditional 
expectations $E_j: M_4\to \iA_j$ are convex combinations of automorphisms
\begin{equation}\label{E:cond}
E_j(A)=\frac{1}{4} \sum_{i=1}^4 U_{ji}^* A U_{ji},
\end{equation}
where $U_{j1}=I$ and $U_{ji}$'s are orthogonal unitaries from $\iA_j'$. 
Since $\iA_5$ is an M-subalgebra, $\iA_5'= \iA_5$. The subalgebras 
$\iA_1',\dots, \iA_4'$ are F-subalgebras generated by the following unitaries:
$$
\begin{array}{ccc}\sigma_1\otimes\sigma_0 & \sigma_2\otimes\sigma_0 
& \sigma_3\otimes\sigma_0, \\\sigma_0\otimes\sigma_1 
& \sigma_1\otimes\sigma_2 & \sigma_1\otimes\sigma_3,\\  
\sigma_2\otimes\sigma_1 & \sigma_0\otimes\sigma_2 
& \sigma_2\otimes\sigma_3,\\ \sigma_3\otimes\sigma_1 
& \sigma_3\otimes\sigma_2 & \sigma_0\otimes\sigma_3.\end{array}
$$
(The above triplets generating $\iA_j$ and $\iA_j'$ ($1 \le j \le 4$) are
Pauli triplets, see \cite{OPSz} for details.) Moreover,
$$
(\Tr A)I= \frac{1}{4}\Big(A+\sum_{j=1}^5 \sum_{k=2}^4 U_{jk}^*AU_{jk}\Big).
$$
The linear mapping (\ref{E:al}) has the concrete form
$$
\alpha(A)=\Big(1-\sum_{i=1}^5\lambda_i \Big) \frac{\Tr A}{4}I+ 
\sum_{i=1}^5 \lambda_i E_i(A),
$$
where the conditional expectations $E_j$ is expressed by the commutant, 
see (\ref{E:cond}). (The condition for complete positivity of $\alpha$
is in Theorem \ref{T:21}.) \qed
\end{pl}

Our main result is the necessary and sufficient condition for the complete 
positivity of mappings like (\ref{E:al}) which can be called generalized 
Pauli channel.
 
\section{Generalized Pauli channels}

Let ${\cal A}$ be a (unital *-) subalgebra of $M_n$. Our aim is to 
describe the conditional expectation onto ${\cal A}$ by means of an 
orthogonal system in the commutant.

Up to unitary equivalence, a subalgebra $\iA$ of $M_n$ can 
be written as 
\[
\iA = \bigoplus_{i=1}^k M_{n_i} \otimes I_{m_i}.
\]
The commutant $\iA'$ in $M_n$ is
\[\iA' = \bigoplus_{i=1}^k I_{n_i} \otimes M_{m_i}.
\]
Let $N=
\sum_{i=1}^k n_i^2$ and let $P_i$ be a minimal central projection 
of $\iA$, that is, $P_i = I_{n_i} \otimes I_{m_i}$.

\begin{prop}\label{cond}
Let $\{U_i\}_{i=1}^N$ be an orthonormal basis of $\iA$. Then the 
completely positive map $F$ from $M_n$ onto $\iA'$ given by
\[
F(X) = \sum_{i=1}^N U_i^* X U_i \qquad (X\in M_n)
\]
is equal to 
\[
F(X) = \bigoplus_{i=1}^k {n_i \over m_i} \Tr_{n_i}(P_i X P_i),
\]
where $\Tr_{n_i}$ is a partial trace from $M_{n_i} \otimes M_{m_i}$ 
onto $M_{m_i}$.

In particular, if all $n_i/ m_i$ are equal, 
then $\frac{n}{{\rm dim}{\iA}}F$ is the trace-preserving conditional expectation from $M_n$ onto $\iA'$.
\end{prop}

\proof
If all $n_i/ m_i$ are equal, then their ratio is equal to ${{\rm dim}\iA \over {n}}$. Therefore it is sufficient to prove the first assertion.

Let $\{e_{ij}^{(l)}\}_{i,j=1}^{n_l} $ and $\{f_{ij}^{(l)}\}_{i,j=1}^{m_l}$ 
be matrix units of $M_{n_l}$ and $M_{m_l}$, respectively. Then $U_i$ is 
written by
\[
U_i = \sum_{l=1}^k \sum_{s,t=1}^{n_l} U_{i,st}^{(l)} e_{st}^{(l)}
\]
for some $U_{i,st}^{(l)} \in {\mathbb C}$.
The operator $W \in M_N$ given in terms of its matrix entries by the formula
\[
W_{i,(l,s,t)}= \sqrt{m_l} U_{i,st}^{(l)}
\]
for $1\le i \le N$, $1\le l \le k$ and $1\le s,t \le n_l$ is unitary.
Indeed, 
$W$ can be considered as the matrix which takes the orthonormal basis
$\{{1\over \sqrt{m_l}}e_{st}^{(l)} \}$ of $\iA$ into the orthonormal basis
$\{ U_i \}$.
Hence we have 
$$
\sum_{i=1}^N \overline{W_{i,(l,s,t)}}W_{i,(l',s',t')}=
\delta_{ll'} \delta_{ss'} \delta_{tt'}.
$$
and therefore
\begin{equation}\label{unitary1.1}
\sum_{i=1}^N \overline{U_{i,st}^{(l)}}U_{i,s't'}^{(l')}=
\delta_{ll'} \delta_{ss'} \delta_{tt'} {1\over m_l}.
\end{equation}

Let $T$ be a partial isometry with $T^*T = e_{s_1s_1}^{(l_1)} 
\otimes f_{t_1t_1}^{(l_1)}$ and $TT^* = e_{s_2s_2}^{(l_2)} 
\otimes f_{t_2t_2}^{(l_2)}$. Then we obtain
\begin{eqnarray*} 
F(T) &=& \sum_{i=1}^N U_i^*TU_i = \sum_{i=1}^N \sum_{p=1}^{n_{l_2}} 
\sum_{q=1}^{n_{l_1}}\overline{U_{i,s_2p}^{(l_2)}}U_{i,s_1q}^{(l_1)}  
e_{ps_2}^{(l_2)} T e_{s_1q}^{(l_1)} \\&=&\delta_{l_1l_2} 
\delta_{s_1s_2}\delta_{pq}\sum_{p=1}^{n_{l_1}} {1\over m_{l_1}} 
e_{ps_1}^{(l_1)} T e_{s_1p}^{(l_1)}
\end{eqnarray*}
by (\ref{unitary1.1}) so that $F$ maps the off-diagonal part to $0$, 
that is, if $l_1 \neq l_2$ then $F(T) =0 $.

Now let $T = e_{s_2s_1}^{(l)} \otimes f_{t_2t_1}^{(l)}$. Then we obtain
\begin{eqnarray*}
F(T) &=& \delta_{s_1s_2} \sum_{p=1}^{n_l} {1\over m_l} e_{pp}^{(l)} 
\otimes f_{t_2t_1}^{(l)}=\delta_{s_1s_2} {1\over m_l} I_{n_l} 
\otimes f_{t_2t_1} \\&=& {n_l \over m_l} {\rm Tr}_{n_l}(T)
\end{eqnarray*}
which shows the first assertion. \qed

The commutant of M-subalgebras and F-subalgebras are again
M-subalgebras and F-subalgebras, and in both types it is possible
to choose an orthogonal basis consisting of unitaries, only.
Thus by an application of the previous proposition,
for such a subalgebra $\iA$, the trace-preserving conditional
expectation is the convex 
combination of automorphisms: 
$$
X \mapsto \frac{1}{{\rm dim}\iA'}\sum_{i=1}^{m}
 U_i'^* X U_i' \qquad (X\in M_n),
$$
where $\{U_i'\}$ is an orthogonal basis of $\iA'$ consisting of unitaries.
Bases consisting of unitaries are important also in quantum state 
teleportation \cite{We}.

\begin{thm}\label{T:A} 
Let $\{U_i: 1 \le i \le m\}$ be an orthonormal system in $M_n$. 
Then the linear mapping
$$
\alpha(A)=\sum_{i=1} \mu_i U_i^* AU_i
$$
is completely positive if and only if
$\mu_i \ge 0$ for every $1\le i \le m$.
\end{thm}

\proof 
If $\mu_i \ge0$ for every $1\le i \le m$,
it is clear that $\alpha$ is completely positive.
To prove the converse, we first show that
$$
\sum_{i,j} W^*E_{ij}W \ot E_{ij}
$$
is a projection if $\Tr WW^* =1 $. This is obviously self-adjoint and
we can compute that it is idempotent:
\begin{eqnarray*}
&& \Big(\sum_{i,j} W^*E_{ij}W \ot E_{ij}\Big)\Big(\sum_{k,l} 
W^*E_{kl}W \ot E_{kl}\Big) \\&=& \sum_{i,j,l} W^*E_{ij}W W^* E_{jl} 
W \ot E_{il} \\ &=& \Tr WW^* \left( \sum_{i,l}  W^*E_{il}W 
\ot E_{il} \right).
\end{eqnarray*}
It follows that 
$$
P_k:=\sum_{i,j} U_k^* E_{ij}U_k \ot E_{ij}
$$
is a projection for every $1 \le k \le m$. To show that they are 
pairwise  orthogonal, we compute the trace of $P_kP_l$:
\begin{eqnarray*}
&&\Tr P_kP_l=\Tr \sum_{i,j,u,v}U_k^* E_{ij}U_kU_l^*E_{uv}U_l 
\ot E_{ij}E_{uv}\cr && \qquad  =\sum_{i,j} \Tr U_k^*E_{ij}
U_k U_l^* E_{ji}U_l=\sum_{i,j} \Tr E_{ij}U_k U_l^* E_{ji}U_lU_k^*.
\end{eqnarray*}
Due to the Lemma \ref{L:1} below this equals $\Tr U_k U_l^* \,
\Tr U_lU_k^*=0$ when $k \ne l$.

The complete positivity implies that
$$
\sum_{i,j}\Big( \sum_k \mu_k U_k^* E_{ij}U_k\ot E_{ij}\Big)
=\sum_{k}\mu_k\Big( \sum_{i,j} U_k^* E_{ij}U_k\ot E_{ij}\Big)
=\sum_{k}\mu_k P_k
$$
is positive, therefore $\mu_k \ge 0$. \qed

\begin{lemma}\label{L:1}
$$
\sum_{i,j} \Tr E_{ij}X E_{ji}Y=(\Tr X)(\Tr Y).
$$
\end{lemma}

\proof
Since both sides are bilinear in the variables $X$ and $Y$, it is 
enough to check the case $X=E_{ab}$ and $Y=E_{cd}$. Simple computation 
gives that left-hand-side is $\delta_{ab}\delta_{cd}$. 
A physicist might make a different proof of the lemma:
$$
\sum_{i,j} \Tr E_{ij}X E_{ji}Y= \sum_{i,j} \Tr |e_i\>\<e_j| X |e_j\>\<e_i|Y
=\sum_{i,j} \<e_j| X |e_j\>\<e_i|Y|e_i\> 
$$
and the right-hand-side is $(\Tr X)(\Tr Y)$. \qed

We also need the next lemma;
the proof can be found in \cite{We}.

\begin{lemma}\label{lemma1}
Let $V_1, V_2, \dots, V_{n^2}$ be matrices in $\Mn$. Then the following 
conditions are equivalent:

1. $\Tr V_i^* V_j= \delta_{ij} \qquad (1 \le i,j \le n^2),$

2. $\sum_{i=1}^{n^2} V_i^*A V_i= (\Tr A)I \mbox{\ for\ every\ } A \in \Mn.$
\end{lemma}

The next result includes important particular cases which are formulated 
afterwards.

\begin{thm}\label{L:**}
Let $\iA_1, 
\dots,\iA_r$ be pairwise complementary subalgebras of $M_n$ such that
their commutants $\iA_1', \dots,\iA_r'$ are pairwise 
complementary as well. Then the trace-preserving 
conditional expectations $E_j: M_n 
\to \iA_j$ can be expressed by the orthonormal 
bases $U_{j1},U_{j2}, \dots, U_{jn(j)} \in \iA_j'$, where $U_{j1}
={1\over \sqrt{n}}I$, via the formula
\begin{equation}\label{eq1}
E_j(A)=\frac{n}{\dim \iA_j'} \sum_{i=1}^{n(j)} U_{ji}^* A U_{ji}
\end{equation}
and the generalized Pauli channel (\ref{E:al}) 
is completely positive if and only if
$$
1+\frac{n^2 \lambda_i}{\dim \iA_i'} \ge \sum_j \lambda_j
$$
for every $1\le i \le r$ and
$$ 
\sum_j \lambda_j\Big(\frac{n^2}{\dim \iA_j'}-1\Big) \ge -1.
$$
\end{thm}

To prove this theorem we prepare the following proposition.

\begin{prop}\label{prop2}
Let $\iA_1$ and $\iA_2$ be complementary subalgebras of $M_n$.
Then $\iA_1'$ and $A_2'$ are complementary if and only if
$\iA_1\iA_2$ linearly spans $M_n$. Moreover, in this case 
the trace-preserving conditional expectation
$E_1: M_n \to \iA_1'$ can be expressed as
\[
E_1(X) = {n \over \dim \iA_1} \sum_{i} U_i^* X U_i \quad (X\in M_n),
\]
where $\{ U_i\}$ is an orthonormal basis of $\iA_1$.
\end{prop}

\proof
Assume $\iA_1'$ and $\iA_2'$ are complementary.
Let $\{U_i'\}$ and $\{V_j'\}$ be orthonormal bases of $\iA_1'$ and $\iA_2'$,
respectively, which consist of scalar multiple of their matrix units.
Then the trace-preserving conditional 
expectations onto $\iA_1$ and $\iA_2$ are given by the linear combinations of 
$U_i'^{*} (\,\cdot\,)U_i'$ and $V_j'^{*} (\,\cdot\,)V_j'$, respectively, 
thanks to Proposition \ref{cond}.

 Since $\iA_1'$ and $\iA_2'$ are complementary subalgebras,
$\{V_j'U_i'\}_{i,j}$ is an orthogonal system.
Moreover the trace is written by the linear combination of  
$ U_i'^*V_j'^* (\,\cdot\,) V_j'U_i'$, because $\iA_1$ and $\iA_2$ are 
complementary subalgebras if and only if the composition of 
 two conditional expectations equals to ${1\over n}\Tr$.
But this shows that $\{V_j'U_i'\}_{i,j}$ linearly spans the whole $M_n$ 
thanks to Lemma \ref{lemma1}. 

Conversely assume $\iA_1 \iA_2$ linearly spans the whole space $M_n$.
Since $\iA_1$ is a subalgebra of $M_n$,
$\iA_1$ can be written as
\[
\iA_1 = \bigotimes_{l=1}^k M_{n_l} \otimes I_{m_l}.
\] 
Let $Q$ be a minimal central projection in $\iA_2$ and let $\{U_i^{(s)}\}$ 
and $\{V_j\}$ are orthonormal bases of $\iA_1$ and $\iA_2$, respectively, 
with the assumption $U_i^{(s)} \in M_{n_s} \otimes I_{m_s}$. Since $\iA_1$ 
and $\iA_2$ are complementary and ${\rm span}\{\iA_1 \iA_2\} = M_n$,
$\{ \sqrt{n} U_i^{(s)} 
V_j \}$ is an orthonormal basis of $M_n$. Therefore by Lemma \ref{lemma1}
and Proposition \ref{cond}, 
we have
$$
\sum_{s,i,j} U_i^{(s)*} V_j^* Q V_j U_i^{(s)} = {1\over n} \Tr Q \cdot I
$$
and
\[\sum_j V_j^*Q V_j = c Q
\]
for some $c >0$. These equations imply, for $1\le s \le k$,
\[
\sum_i U_i^{(s)*} Q  U_i^{(s)} = {\Tr Q \over cn} P_s,
\]
where $P_s$ is a central projection $I_{n_s}\otimes I_{m_s}$.
Now we take the trace to the above equation. Then we have
\begin{eqnarray*}
\Tr \left(\sum_i^{n_s^2} U_i^{(s)*} Q  U_i^{(s)}\right)
= \sum_{i}^{n_s^2} {1\over n} \Tr \left( U_i^{(s)}U_i^{(s)*}\right) 
\Tr Q={n_s^2 \over n} \Tr Q
\end{eqnarray*}
and
\[
{\Tr Q \over cn}\Tr P_s = {\Tr Q \over cn} n_s m_s
\]
so that
\[
{n_s \over m_s} = {1 \over c}.
\]
Hence ${n_s / m_s}$ is equal to ${1/ c}={\dim \iA_1 \over n}$ for all 
$1\le s \le k$ and so \[E_1 = {n \over \dim \iA_1 } \sum_{s,i} U_i^{(s)*} 
(\, \cdot \,) U_i^{(s)}\]is the trace-preserving 
conditional expectation onto $\iA_1'$ 
by Proposition \ref{cond}. Similarly, 

\[
E_2 =  {n \over \dim\iA_2}\sum_{j} V_j^{*} (\, \cdot \,) V_j.
\]
is the trace-preserving conditional expectation onto $\iA_2'$. Since 
\[
\sum_{s,i,
j} U_i^{(s)*} V_j^* (\, \cdot \,) V_j U_i^{(s)} 
=\sum_{s,i,j} 
V_j^* U_i^{(s)*} (\, \cdot \,) U_i^{(s)} V_j
\]
is the normalized trace 
on $M_n$ by Lemma \ref{lemma1}, we obtain
${n^2 \over \dim \iA_1 \dim \iA_2}=1$ and so the composition  
$E_1 \circ E_2$ equals to ${1\over n} \Tr$.\qed

{\it Proof of Theorem \ref{L:**}.}
The first assertion is already proven in the above proposition.
Due to the Lemma \ref{lemma1}, we have
$$
(\Tr A)I= {A\over n} +\sum_{j=0}^n \sum_{k=2}^n U_{jk}^*AU_{jk}+
\sum_{t=1}^\ell W_t^* A W_t,
$$
where orthonormal system $W_t$ extend the orthonormal system $U_{jk}$ 
to a complete system in the linear space $\Mn$. In formula (\ref{E:al}) 
we use this expression for 
$(\Tr A)I$ and the assumed decomposition of the conditional expectations. 
So in the expansion of $\alpha(A)$ the coefficient of $U_{jk}^*AU_{jk}$ is
$$
\frac{1}{n}\Big(1-\sum_i \lambda_i\Big)+ \frac{n\lambda_j}{\dim \iA_j'}
$$
and the coefficient of ${A\over n} = 
\left({1\over \sqrt{n}}I\right) A\left({1\over \sqrt{n}}I\right)$ is
$$
\frac{1}{n}\Big(1-\sum_j \lambda_j\Big)+ \sum_j 
\frac{n\lambda_j}{\dim\iA_j'}.
$$
Theorem \ref{T:A} tells us that completely positivity
holds if and only if both are positive. \qed

\begin{cor}\label{T:2} 
Assume that $\Mn$ contains pairwise complementary M-subalgebras
$\iA_1, \dots,\iA_r$. Then the generalized Pauli channel is completely 
positive if and only if
$$
1+ n \lambda_i \ge\sum_j \lambda_j \ge -\frac{1}{n-1}
$$
for every $1 \le i \le r$.
\end{cor}
This result appeared also in \cite{NR}.

\section{Bipartite channels}

In this section we consider subalgebras of $\Mn \ot \Mn$. A subalgebra
isomorphic to $\Mn$ will be called F-subalgebra. An M-subalgebra is a 
maximal Abelian subalgebra. Both kinds of subalgebras are subspaces of 
dimension $n^2$.

\begin{thm}\label{T:3} 
Assume that $\iA_1$ and $\iA_2$ are F- or M-subalgebras of $\Mn \ot \Mn$. 
If they are complementary, 
then the commutants $\iA_1'$ and $\iA_2'$ are complementary as well.
\end{thm}
 
\proof
Since both kinds of subalgebras are subspaces of dimension $n^2$,
the dimension of $\iA_1 \iA_2$ is $n^4$ so that $\iA_1 \iA_2 = M_n \ot M_n$.
Therefore the commutants $\iA_1'$ 
and $\iA_2'$ are complementary by Proposition \ref{prop2}.
\qed


\begin{thm}\label{T:21}
Assume that $\Mn \ot \Mn$ is decomposed to pairwise complementary F- and 
M-subalgebras $\iA_i$ ($1 \le i \le n^2+1$). The trace-preserving 
conditional expectation of $\Mn \ot \Mn$ onto $\iA_i$ is denoted by $E_i$. 
The linear trace-preserving mapping acting as
$$
\alpha(B)=\sum_{i=1}^{n^2+1} \lambda_i E_i(B) \qquad (B \in \Mn \ot 
\Mn, \,\Tr B=0)
$$
is completely positive if and only if
$$
1+ n^2 \lambda_i \ge\sum_j \lambda_j \ge -\frac{1}{n^2-1}
$$
for every $1 \le i \le n^2+1$.
\end{thm}

\proof
Theorem \ref{T:3} allows to use Theorem \ref{L:**} and the result follows.
\qed

The theorem can be applied in Example \ref{Ex:2}. Note that decompositions
of $M_2 \ot M_2$ into F- and M-subalgebras are discussed in \cite{m4},
while decomposition of $M_n \ot M_n$ into F-subalgebras is constructed
in \cite{Ohno} if $n=p^k$ with a prime number $p>2$. 

\bigskip\bigskip{\bf Acknowledgement.} The authors thank to Professor 
Tsuyoshi Ando for communication and to the project of JSPS and Hungarian 
Academy of Sciences for support.


\begin{thebibliography}{99}

\bibitem{BFS}C. H. Bennett, C. A. Fuchs and J. A. Smolin, Entanglement-enhanced classical communication on a noisy quantum 
channel, pp. 79-88 in {\it Quantum Communication, Computing, and Measurement} (eds. O. Hirota, A.S. Holevo, and C.M. Caves) Plenum, New York, 1997. 

\bibitem{Choi}
M-D. Choi, Completely positive linear maps on complex matrices, Linear 
Alg. Appl. {\bf 10}(1975), 285--290.

\bibitem{Fuji}
A. Fujiwara and H. Imai, Quantum parameter estimation of a generalized Pauli 
channel, J. Phys. A: Math. Gen. {\bf 36}(2003), 8093-8103.

\bibitem{King}C. King, Additivity for unital qubit channels, J. Math. Phys. 
{\bf 43}(2002), 4641.

\bibitem{NR}
M. Nathanson and M.B. Ruskai, Pauli diagonal channels constant on axes,
J. Phys. A: Math. Theor. {\bf 40}(2007),  8171--8204.

\bibitem{Ohno}
H. Ohno, Quasi-orthogonal subalgebras of matrix algebras, Linear Alg. Appl. 
{\bf 429}(2008), 2146--2158.

\bibitem{OPSz}
H. Ohno, D. Petz and A. Sz\'ant\'o, Quasi-orthogonal subalgebras of 
$4 \times 4$ matrices, Linear Alg. Appl. {\bf 425}(2007), 109--118.

\bibitem{PDcomp}
D. Petz, Complementarity in quantum systems, Rep. Math. Phys. 
{\bf 59}(2007), 209--224.

\bibitem{m4}
D. Petz, A. Sz\'ant\'o and M. Weiner, Complementarity and the algebraic 
structure of 4-level quantum systems, to be published in Infin. Dimens. 
Anal. Quantum Probab. Relat. Top.

\bibitem{pd2}
D. Petz, {\it Quantum Information Theory and Quantum Statistics}, Springer, 
Berlin, Heidelberg, 2008.

\bibitem{MUB}
A.O. Pittenger and M.H. Rubin, Mutually unbiased bases, generalized spin matrices and separability,  Linear Alg. Appl.  {\bf 390}(2004), 255--278.

\bibitem{Zych} 
W. Tadej and K. Zyczkowski, A concise guide to complex Hadamard matrices, 
Open Syst. Inf. Dyn. {\bf 13}(2006), 133-177. 

\bibitem{We}
R.F. Werner, All teleportation and dense coding schemes,  J. Phys. 
{\bf A34}(2001), 7081--7094.

\end{thebibliography}
\end{document}